\documentclass{emulateapj}
\usepackage{amssymb,amsmath,graphicx,longtable,subfigure,wrapfig}
\usepackage[colorlinks,linkcolor=blue,anchorcolor=green,citecolor=blue]{hyperref}

\begin{document}

\def\func#1{\mathop{\rm #1}\nolimits}
\def\unit#1{\mathord{\thinspace\rm #1}}

\title{Solving the $^{56}$Ni puzzle of magnetar-powered broad-lined type Ic
supernovae}
\author{Ling-Jun Wang\altaffilmark{1}, Yan-Hui Han\altaffilmark{1}, Dong Xu%
\altaffilmark{1}, Shan-Qin Wang\altaffilmark{2,3}, Zi-Gao Dai%
\altaffilmark{2,3}, Xue-Feng Wu\altaffilmark{4,5}, Jian-Yan Wei%
\altaffilmark{1}}

\begin{abstract}
Broad-lined type Ic supernovae (SNe Ic-BL) are of great importance because
their association with long-duration gamma-ray bursts (LGRBs) holds the key
to deciphering the central engine of LGRBs, which refrains from being
unveiled despite decades of investigation. Among the two popularly
hypothesized types of central engine, i.e., black holes and strongly
magnetized neutron stars (magnetars), there is mounting evidence that the
central engine of GRB-associated SNe (GRB-SNe) is rapidly rotating
magnetars. Theoretical analysis also suggests that magnetars could be the
central engine of SNe Ic-BL. What puzzled the researchers is the fact that
light curve modeling indicates that as much as $0.2-0.5$ solar mass of $%
^{56} $Ni was synthesized during the explosion of the SNe Ic-BL, which is
unfortunately in direct conflict with current state-of-the-art understanding
of magnetar-powered $^{56}$Ni synthesis. Here we propose a dynamic model of
magnetar-powered SNe to take into account the acceleration of the ejecta by
the magnetar, as well as the thermalization of the injected energy. Assuming
that the SN kinetic energy comes exclusively from the magnetar acceleration,
we find that although a major fraction of the rotational energy of the
magnetar is to accelerate the SNe ejecta, a tiny fraction of this energy
deposited as thermal energy of the ejecta is enough to reduce the needed $%
^{56}$Ni to 0.06 solar mass for both SNe 1997ef and 2007ru. We therefore
suggest that magnetars could power SNe Ic-BL both in aspects of energetics
and of $^{56}$Ni synthesis.
\end{abstract}

\keywords{stars: neutron --- supernovae: general --- supernovae: individual
(SN 1997ef, SN 2007ru)}

\affil{\altaffilmark{1}Key Laboratory of Space Astronomy and Technology,
National Astronomical Observatories,
Chinese Academy of Sciences, Beijing 100012, China; wanglj@nao.cas.cn, wjy@nao.cas.cn}

\affil{\altaffilmark{2}School of Astronomy and Space Science, Nanjing
University, Nanjing 210093, China; dzg@nju.edu.cn}

\affil{\altaffilmark{3}Key Laboratory of
Modern Astronomy and Astrophysics (Nanjing University),
Ministry of Education, Nanjing 210093, China}

\affil{\altaffilmark{4}Purple Mountain Observatory, Chinese Academy of
Sciences, Nanjing, 210008, China}

\affil{\altaffilmark{5}Joint Center for Particle Nuclear Physics and Cosmology of Purple Mountain Observatory-Nanjing University,
Chinese Academy of Sciences, Nanjing 210008, China}

\section{Introduction}

It is widely accepted that the death of massive stars should trigger
core-collapse supernovae \citep[CCSNe;][]{Bethe90,Janka12} that can be
classified as types IIP, IIL, IIn, IIb, Ib and Ic \citep{Filippenko97}. In
the last two decades, some SNe Ic having broader P-Cygni profiles and
absorption troughs than normal SNe Ic were confirmed and nominated as
\textquotedblleft broad-lined SNe" \citep[SNe Ic-BL;][]{Woosley06}.

Some SNe Ic-BL are associated with gamma-ray bursts (GRBs) or X-ray flashes %
\citep[XRFs;][]{Woosley06,Cano16}. The association of LGRBs with SNe Ic-BL
provides a unique channel to study the central engine of GRBs. Before the
discovery of SNe Ic-BL, the majority of conventional SNe has a kinetic
energy of $\sim 10^{51}\unit{erg}$, which is generally attributed to
neutrino energy deposition \citep{Woosley02,Janka12}. The huge amount of
kinetic energy of SNe Ic-BL, $\sim 10^{52}\unit{erg}$, poses an immediate
challenge to this canonical SN picture.

One way to generate such a tremendous kinetic energy is to assume that the
explosion remnant is a rapidly rotating magnetar %
\citep{Wheeler00,Thompson04,Wang16b}, whose rotational energy is converted
as the kinetic energy of SNe Ic-BL. Indeed, it is found that the kinetic
energies of SNe Ic-BL associated with LGRBs are clustered at $10^{52}\unit{%
erg}$ with an upper limit of $\sim 2\times 10^{52}\unit{erg}$ %
\citep{Mazzali14}, namely the maximum rotational energy of magnetars. This
is a strong clue that GRB-SNe are powered by millisecond magnetars. In
addition, the light curve of SN 2011kl associated with the ultra-long GRB
111209A suggests the existence of magnetar because $^{56}$Ni is inadequate
to reproduce the observational data \citep{Greiner15}.

Light curve modeling of SNe Ic-BL indicates the synthesis of $^{56}$Ni as
massive as $M_{\mathrm{Ni}}=0.2-0.5M_{\odot }$, where $M_{\odot }$ is the
solar mass. However, theoretical studies found that it is very difficult to
synthesize $0.2M_{\odot }$ of $^{56}$Ni by a millisecond magnetar with
parameters given in the literature \citep{Nishimura15,Suwa15}. This conflict
is a big concern to accept the hypothesis that SNe Ic-BL are powered by
magnetars.

In arriving at the conclusion that SNe Ic-BL must have synthesized as
massive as $0.2-0.5M_{\odot }$ of $^{56}$Ni when modeling the SN light
curves, one usually assumes that the SN thermal energy comes exclusively
from the thermalization of the gamma-rays from the decay of $^{56}$Ni and $%
^{56}$Co. This assumption is correct if the thermalization of the (assumed)
magnetar spin-down power can be neglected compared to the energy deposition
from the decay of $^{56}$Ni and $^{56}$Co, as in the case of ordinary SNe Ic.

In the magnetar model for optical transients, it is well known that the
contribution of magnetar to the SN thermal emission dominates over other
(possible) energy sources in the case of superluminous SNe 
\citep[SLSNe;][]{Kasen10,Woosley10,Chatzopoulos12,Chatzopoulos13,
Inserra13,Nicholl14,Metzger15,WangLiu16,WangWang15,Dai16,Kashiyama16}. Even
for luminous SNe, whose luminosities lie between normal SNe and SLSNe, the
contribution from magnetar dominates during the early times after SN
explosion \citep{WangWang2015b}.

SNe Ic-BL, though very energetic in aspects of their kinetic energy, are
much less luminous than SLSNe and luminous SNe and are comparable to or
slightly luminous than normal SNe Ic. Just for this reason it is believed
that the luminosities of SNe Ic-BL are the result of $^{56}$Ni heating. At
first glance this view seems correct because it is suggested that the
spin-down timescales of the magnetar powering the SNe Ic-BL are very short
so that the rotational energy of the magnetar is exhausted in accelerating
the SN ejecta and little is left to heat the SN \citep{Wang16b}. SLSNe
instead are so luminous because the spin-down timescales of the magnetars
are much longer so that a significant fraction of their rotational energy is
utilized to heat the SNe \citep{Wang16b}.

In view of the moderate luminosity of SNe Ic-BL and the fact that the
rotational energy of the magnetars cannot completely deposit as the kinetic
energy of the ejecta, we suspect that the magnetars could contribute to the
luminosity of SNe Ic-BL significantly and hence reduce the needed $^{56}$Ni.
If this is the case, the conflict of high mass $^{56}$Ni in modeling the SN
light curves and the low yield of $^{56}$Ni produced by the magnetar-driven
shock \citep{Nishimura15,Suwa15} can be solved. This is the motivation for
the work presented here. To this end we present our model in Section \ref%
{sec:model} and then apply it to two carefully selected SNe Ic-BL \ in
Section \ref{sec:results}. Implications of our findings are discussed in
Section \ref{sec:dis}.

\section{The Dynamic Model}

\label{sec:model}

To determine the fraction of the rotational energy of the magnetar that
deposits as the thermal energy of the SN, we need a model to deal with the
acceleration and heating of the SN ejecta by the magnetar spin-down power in
a self-consistent way. The kinetic energy of the SN is given by %
\citep{Arnett82}%
\begin{equation}
E_{\mathrm{SN}}=\frac{3}{10}M_{\mathrm{ej}}v_{\mathrm{sc}}^{2},
\end{equation}%
where $M_{\mathrm{ej}}$ is the ejecta mass, the scale velocity $v_{\mathrm{sc%
}}$ evolves according to \citep{Wang16b}%
\begin{equation}
v_{\mathrm{sc}}=\left[ \left( \frac{5}{3}\right) \frac{2\left( E_{\mathrm{SN}%
,0}+E_{K,\mathrm{inp}}\right) }{M_{\mathrm{ej}}}\right] ^{1/2}.
\label{eq:v_t-evolution}
\end{equation}%
Here $E_{\mathrm{SN},0}$ is the initial kinetic energy of the SN and the
magnetar's kinetic energy input $E_{K,\mathrm{inp}}$ is given by the energy
conservation condition%
\begin{equation}
\frac{dE_{K,\mathrm{inp}}}{dt}=L_{K}-L,  \label{eq:E_K_evolve}
\end{equation}%
where $L$ is the SN luminosity. In Equations $\left( \ref{eq:v_t-evolution}%
\right) $ and $\left( \ref{eq:E_K_evolve}\right) $ we neglect internal
energy because its effect is to change the effective mass of the ejecta,
which is negligibly small compared to the ejecta mass. The kinetic energy
input rate from the magnetar, $L_{K}$, is given by%
\begin{equation}
L_{K}=L_{\mathrm{mag}}\left( t\right) \left( 1-e^{-\tau _{\gamma ,\mathrm{mag%
}}}\right) ,  \label{eq:K_input_rate_mag}
\end{equation}%
where 
\begin{equation}
L_{\mathrm{mag}}\left( t\right) =\frac{E_{\mathrm{sd}}}{\tau _{\mathrm{sd}%
}\left( 1+t/\tau _{\mathrm{sd}}\right) ^{2}}
\end{equation}%
is the spin-down power of the magnetar. Here $\tau _{\mathrm{sd}}=2.3\unit{%
days}R_{\ast ,6}^{-6}B_{p,14}^{-2}P_{0,-3}^{2}$ is the spin-down timescale
of the magnetar, $E_{\mathrm{sd}}=L_{\mathrm{sd},0}\tau _{\mathrm{sd}}$, $L_{%
\mathrm{sd},0}=10^{47}\unit{erg}\unit{s}^{-1}P_{0,-3}^{-4}B_{p,14}^{2}R_{%
\ast ,6}^{6}$ is the spin-down luminosity of the magnetar. Here the
convention $Q=10^{n}Q_{n}$ is adopted in the c.g.s. units. $R_{\ast }$, $%
P_{0}$, $B_{p}$ are the radius, initial rotational period, magnetic dipole
field of the magnetar, respectively. $\tau _{\gamma ,\mathrm{mag}}$ is the
optical depth of the ejecta to gamma-rays emitted by the spinning down
magnetar. The factor $\left( 1-e^{-\tau _{\gamma ,\mathrm{mag}}}\right) $ in
Equation $\left( \ref{eq:K_input_rate_mag}\right) $ is to take account for
the hard photon leakage from magnetar (\citealt{WangWang15}, see also %
\citealt{Chen15}). Because the energy spectra of radioactive decay photons
and magnetar spin-down photons are different, two $\kappa _{\gamma }$'s,
namely the opacity to magnetar spin-down photons $\kappa _{\gamma ,\mathrm{%
mag}}$ and to radioactive decay photons $\kappa _{\gamma ,\mathrm{decay}}$
are used here. In this paper therefore three opacities are used, i.e. the
opacity to visible photons $\kappa $, the opacities to the $\gamma $-ray
photons from magnetars and radioactive decay photons, $\kappa _{\gamma ,%
\mathrm{mag}}$ and $\kappa _{\gamma ,\mathrm{decay}}$, respectively. The
introduction of above equations is the key to determining the fraction of
the rotational energy of the magnetar that deposits as the thermal energy of
the SN.

The SN luminosity is given by \citep{Arnett82}%
\begin{equation}
L=\frac{E_{\mathrm{th}}\left( 0\right) }{\tau _{0}}\phi \left( t\right) ,
\label{eq:SN-L-no-recede}
\end{equation}%
where $E_{\mathrm{th}}\left( 0\right) $ is the initial thermal energy of the
SN, $\phi \left( t\right) $ evolves according to%
\begin{equation}
\dot{\phi}=\frac{R\left( t\right) }{R\left( 0\right) }\left[ \frac{L_{%
\mathrm{inp}}\left( t\right) }{E_{\mathrm{th}}\left( 0\right) }-\frac{\phi }{%
\tau _{0}}\right] .  \label{eq:SN-phi-dot-no-recede}
\end{equation}%
The diffusion timescale $\tau _{0}$ is%
\begin{equation}
\tau _{0}=\frac{\kappa M_{\mathrm{ej}}}{\beta cR\left( 0\right) },
\label{eq:diffusion-time}
\end{equation}%
where $\beta \simeq 13.8$, and $R\left( t\right) $ is the SN radius at time $%
t$. The energy input $L_{\mathrm{inp}}\left( t\right) $ includes two
sources, i.e. $^{56}$Ni (plus $^{56}$Co) decay energy and magnetar spin-down
power 
\begin{equation}
L_{\mathrm{inp}}\left( t\right) =L_{\mathrm{mag}}\left( t\right) \left(
1-e^{-\tau _{\gamma ,\mathrm{mag}}}\right) +L_{\mathrm{Ni}}\left( t\right)
\left( 1-e^{-\tau _{\gamma ,\mathrm{decay}}}\right)
\end{equation}%
with%
\begin{equation}
L_{\mathrm{Ni}}\left( t\right) =M_{\mathrm{Ni}}\left[ \left( \epsilon _{%
\mathrm{Ni}}-\epsilon _{\mathrm{Co}}\right) e^{-t/\tau _{\mathrm{Ni}%
}}+\epsilon _{\mathrm{Co}}e^{-t/\tau _{\mathrm{Co}}}\right] ,
\end{equation}%
where $\epsilon _{\mathrm{Ni}}=3.9\times 10^{10}\unit{erg}\unit{g}^{-1}\unit{%
s}^{-1}$, $\epsilon _{\mathrm{Co}}=6.78\times 10^{9}\unit{erg}\unit{g}^{-1}%
\unit{s}^{-1}$, $\tau _{\mathrm{Ni}}$ and $\tau _{\mathrm{Co}}$ are the
lifetime of $^{56}$Ni and $^{56}$Co, respectively. In deriving Equations $%
\left( \ref{eq:SN-L-no-recede}\right) $ and $\left( \ref%
{eq:SN-phi-dot-no-recede}\right) $ we assume that the injected energy is
trapped as internal energy.\footnote{%
This is an approximation because the injected energy should be divided into
internal energy of random motion and kinetic energy of directed motion. To
accurately determine how much fraction of the injected energy goes into
internal energy, one should carry out more elaborated calculation to take
account of the scattering of photons by electrons. Numerical simulations
indicate that a strong shock deposits its energy equally into directed
kinetic energy and random internal energy. Here we just assume that the
equations derived since the first formulation of the Arnett model is
reasonably correct so that we can utilize their result.}

In this model, the scale velocity $v_{\mathrm{sc}}$ is not a constant so
that Equation $\left( \ref{eq:SN-phi-dot-no-recede}\right) $ cannot be
expressed as an integration equation, as in the usual Arnett model. What we
can expect from Equation $\left( \ref{eq:E_K_evolve}\right) $ is the rapid
acceleration of the ejecta during early times when $L_{K}>L$. To efficiently
convert the rotational energy of the magnetar into SN kinetic energy, the
magnetar must deposit its rotational energy when the ejecta is very compact
so that its optical depth is essentially infinite. This condition can be
fulfilled only if the spin-down timescale is very short. On the other hand,
to make a bright SN, e.g. an SLSN, the magnetar must retain its rotational
energy for a much long time before the SN ejecta expand to a very large
distance. In this case, the ejecta gain little kinetic energy.

\section{Sample selection and Results}

\label{sec:results}

In this work we would like to avoid the SNe that show clear aspheric
expansion because the above analytic model assumes a homologous and
spherical expansion. In line with this criterion, we exclude GRB-SNe in this
work because any SN associated with a GRB is accompanied by a relativistic
jet and is therefore aspheric. Some SNe Ic-BL not associated with GRBs are
also aspheric because a non-negligible fraction of their ejecta is moving at
relativistic speed, e.g. SN 2009bb \citep{Pignata11} and SN 2012ap %
\citep{Milisavljevic15}.

With the above criterion borne in mind, we searched the literature and found
that there are currently about ten SNe Ic-BL that are not associated with
GRBs and also do not show evidence for relativistic outflow.

As will be clear, assuming SNe Ic-BL are powered by magnetars, it is found
that the early-time light curves of SNe Ic-BL are mainly determined by the
parameters of magnetars, whereas the late-time light curves are determined
dominantly by the mass of $^{56}$Ni. To unambiguously evaluate the mass of $%
^{56}$Ni, we should select the SNe Ic-BL such that their observational data
extend at least to $>100\unit{days}$. By doing so we are sampling the decay
tail of $^{56}$Co because the lifetime of $^{56}$Co is $111.3\unit{days}$.
To accurately determine the parameters of the magnetars that powers the SNe
Ic-BL, there should be a good sampling in the observational data before the
maximum of the SN light curve.

\begin{figure}[tbph]
\centering\includegraphics[width=0.4\textwidth,angle=0]{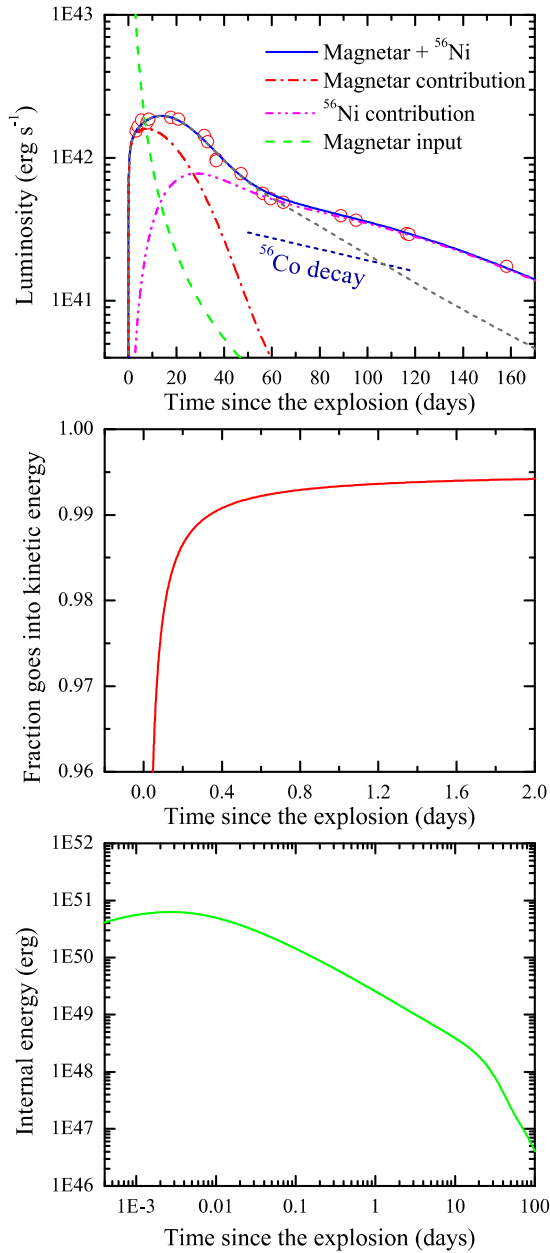}
\caption{The light curve (top), the fraction of magnetar rotational energy
deposited as the kinetic energy of SN 1997ef (middle), and the internal
energy of the SN (bottom). In the top panel the solid line is the light
curve produced by taking account for the contribution from both magnetar and 
$^{56}$Ni, while the dot-dashed line is the light curve by setting the mass
of $^{56}$Ni zero while other parameters are the same as that of the solid
line. The dot-dot-dashed line is the difference between solid line and the
dot-dashed line. The dark short-dashed line is the light curve with $\protect%
\kappa _{\protect\gamma ,\mathrm{decay}}=0.027\unit{cm}^{2}\unit{g}^{-1}$.
Please note that the abscissa time scales in these panels are quite
different and the last panel is in logarithmic scale. The data points are
taken from \protect\cite{Iwamoto00}.}
\label{fig:1997ef}
\end{figure}

\begin{figure}[tbph]
\centering\includegraphics[width=0.4\textwidth,angle=0]{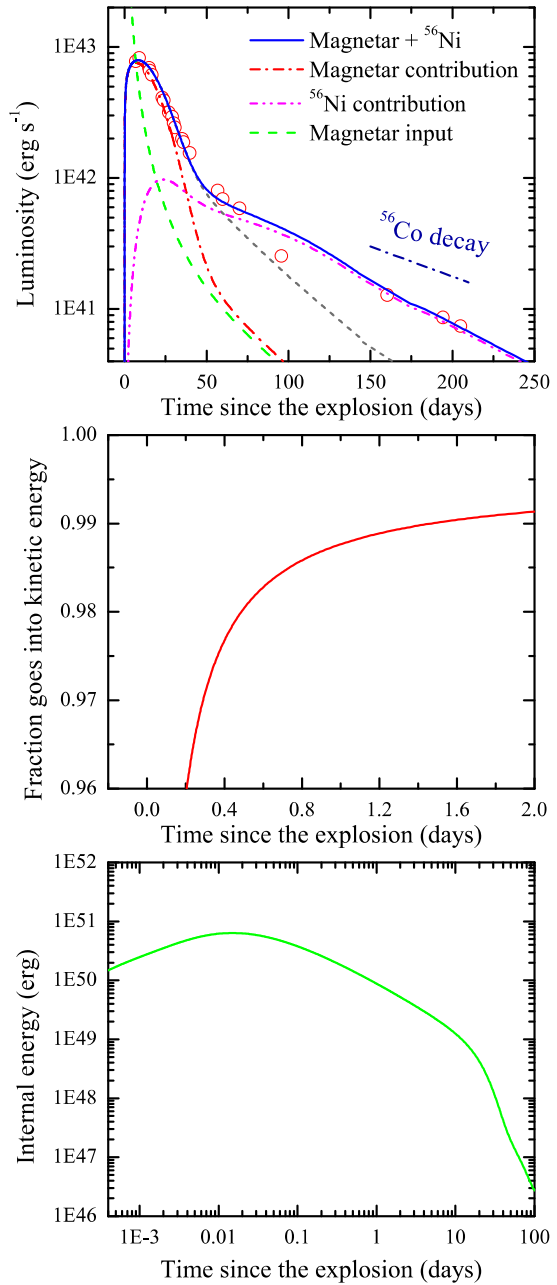}
\caption{The same as Figure \protect\ref{fig:1997ef} but for SN 2007ru. The
data points are taken from \protect\cite{Sahu09}.}
\label{fig:2007ru}
\end{figure}

With these two additional criteria we find we are left with two SNe Ic-BL,
namely SNe 1997ef \citep{Iwamoto00} and 2007ru \citep{Sahu09}. In the light
curve modeling of SNe Ic, the opacity $\kappa _{\gamma ,\mathrm{decay}}$ to
radioactive decay photons usually takes the value $\kappa _{\gamma ,\mathrm{%
decay}}\sim 0.025-0.027\unit{cm}^{2}\unit{g}^{-1}$ 
\citep[e.g.,][and
references therein]{WangWang2015b}. In Figures \ref{fig:1997ef} and \ref%
{fig:2007ru}, we show the light curves with $\kappa _{\gamma ,\mathrm{decay}%
}=0.027\unit{cm}^{2}\unit{g}^{-1}$, which evidently fail to reproduce the
light curves. One common feature of these two SNe is that their linear decay
phase is consistent with nearly full trapping of $^{56}$Co. Given this fact,
the $^{56}$Ni mass can be accurately determined by modeling the late-time
light curve of the SN because the contribution of magnetar at late times is
negligible for SNe Ic-BL \citep{Wang16b}. In the top panels of Figures \ref%
{fig:1997ef} and \ref{fig:2007ru} the solid lines are the synthesized light
curves assuming full trapping. The full trapping is not rare for SNe Ic
given that SN 2007bi also has a linear decay phase that is consistent with
full trapping \citep{Gal-Yam09}. This could indicate that these SNe have
some nontrivial density structure.

The ejecta mass $M_{\mathrm{ej}}$ can be determined by equating the light
curve rising time to the following effective diffusion timescale %
\citep{Arnett82}%
\begin{equation}
\tau _{m}=\left( \frac{2\kappa M_{\mathrm{ej}}}{\beta cv_{\mathrm{ph}}}%
\right) ^{1/2},  \label{eq:tau_m-SN}
\end{equation}%
where $v_{\mathrm{ph}}\approx v_{\mathrm{sc}}$ is the photospheric velocity
of the SN. The optical opacity is fixed at $\kappa =0.1\unit{cm}^{2}\unit{g}%
^{-1}$ in this work. However, one should not equate $\tau _{m}$ with the
apparent rising time of SNe Ic-BL because their light curve cannot be
reproduced by pure $^{56}$Ni heating. Instead, one should isolate the $^{56}$%
Ni contribution from the apparent light curve, as demonstrated in Figures %
\ref{fig:1997ef} and \ref{fig:2007ru} and equate $\tau _{m}$ to the rising
time of $^{56}$Ni contribution. One may be confused why we should
discriminate the rising times from $^{56}$Ni contribution and magnetar
contribution because Equations $\left( \ref{eq:SN-L-no-recede}\right) $ and $%
\left( \ref{eq:SN-phi-dot-no-recede}\right) $ do not care if the energy
input comes from the magnetar or the radioactive decay. Actually in deriving
Equation $\left( \ref{eq:tau_m-SN}\right) $ we implicitly assume that the
energy release timescale $\tau _{\mathrm{release}}$ is comparable to or
longer than the ejecta expansion timescale $\tau _{\exp }$, i.e. $\tau _{%
\mathrm{release}}\gtrsim \tau _{\exp }$. The reason that at time $\tau _{m}$
the luminosity reaches its peak is as follows. Physical intuition tells us
that the SN reaches peak luminosity when most of the available energy has
the right time to diffuse out of the SN and at the same time the SN expands
to a considerable distance so that its emitting surface is sufficiently
large. Then equating the diffusion timescale $\left( \ref{eq:diffusion-time}%
\right) $ to the expansion timescale $\tau _{\exp }=R\left( 0\right) /v_{%
\mathrm{sc}}$ immediately leads to the effective diffusion timescale $\left( %
\ref{eq:tau_m-SN}\right) $.\footnote{%
More elaborated calculation gives the factor 2 in Equation $\left( \ref%
{eq:tau_m-SN}\right) $.} However, the condition $\tau _{\mathrm{release}%
}\gtrsim \tau _{\exp }$ is true for $^{56}$Ni decay, but not for magnetar
input, which has a release timescale $\tau _{\mathrm{sd}}\sim 10^{-3}\unit{%
days}$ for both SNe 1997ef and 2007ru. Because the magnetar releases its
energy in such a short time, the SN ejecta have no time to expand. As a
result, its peak luminosity occurs at the time when the magnetar release
most of its energy. In this aspect, the magnetar-powered SNe are more or
less similar to the explosive energy release found in some type II SNe where
the peak luminosity occurs at the time when the SNe explode \citep{Arnett80}%
. This analysis indicates that, depending on the relative relation of the
two timescales, $\tau _{\mathrm{release}}$ and $\tau _{\exp }$, the
magnetar-powered SN light curve could be similar to type I SNe or type II
SNe.

Given the ejecta mass, the kinetic energy of the SN can be evaluated. Here
we adopt the simple but quite plausible assumption that the kinetic energy
of the SN is exclusively injected by the rapidly spinning magnetar.
Consequently, the initial rotational period $P_{0}$ of the magnetar can be
determined. The magnetic dipole field of the magnetar, on the other hand,
could be determined by modeling the early-time light curve. In this way, the
four parameters in this model can all be tightly constrained.

Figures \ref{fig:1997ef} and \ref{fig:2007ru} show the results of the
analytical model with best-fit parameters listed in Table \ref{tbl:para}. In
these fits, we adopt the widely used value $\kappa =0.1\unit{cm}^{2}\unit{g}%
^{-1}$ so that there are effectively four free parameters in this model.
From Figures \ref{fig:1997ef} and \ref{fig:2007ru} it is clear that the
contribution to the light curves from magnetars at late times is negligible.

\begin{table}[tbph]
\caption{Best-fit parameters for SNe 1997ef and 2007ru. In these fits, we
fix $\protect\kappa =0.1\unit{cm}^{2}\unit{g}^{-1}$.}
\label{tbl:para}
\begin{center}
\begin{tabular}{ccccc}
\hline
SN & $M_{\mathrm{ej}}$ & $M_{\mathrm{Ni}}$ & $B_{p}$ & $P_{0}$ \\ 
\multicolumn{1}{l}{} & $\left( M_{\odot }\right) $ & $\left( M_{\odot
}\right) $ & $\left( 10^{16}\unit{G}\right) $ & $\left( \unit{ms}\right) $
\\ \hline
\multicolumn{1}{l}{SN 1997ef} & \multicolumn{1}{l}{$5.9$} & $0.056$ & $1.24$
& $2.25$ \\ 
\multicolumn{1}{l}{SN 2007ru} & \multicolumn{1}{l}{$4.43$} & $0.061$ & $0.62$
& $2.30$ \\ \hline
\end{tabular}%
\end{center}
\end{table}

Comparison of Figure \ref{fig:1997ef} with the $^{56}$Ni-powered light curve %
\citep{Iwamoto00} immediately shows the superior fitting quality of the
light curve in Figure \ref{fig:1997ef}. It is worthy of mentioning that the
above analytical model, as a direct extension of the Arnett model %
\citep{Arnett82}, is quite good at reproducing the light curve of a purely $%
^{56}$Ni-powered SN \citep{Arnett89}. When applying the pure $^{56}$Ni model
to SNe Ic-BL, the mass of $^{56}$Ni is determined by the peak luminosity of
the SN. But the $^{56}$Co tail modeling usually requires a much lower mass
of $^{56}$Ni. Hence an inconsistency appears. In other words, the poor
fitting quality of the pure $^{56}$Ni model is intrinsic for SNe Ic-BL.

Figures \ref{fig:1997ef} and \ref{fig:2007ru} indicate that the (assumed)
magnetars contribute dominantly to the early peaks of the light curves. This
is the key that the actually needed $^{56}$Ni, $M_{\mathrm{Ni}}\sim
0.06M_{\odot }$ for both SNe 1997ef and 2007ru, is much lower than a pure $%
^{56}$Ni model. It is also evident from these figures that the late-time
luminosities of these SNe dominantly come from the decay energy of $^{56}$Ni
and $^{56}$Co.

Table \ref{tbl:para} shows that the ejecta masses $M_{\mathrm{ej}}$ are
different from that given by the pure-$^{56}$Ni model, which favors the
ejecta masses $M_{\mathrm{ej}}=7.6M_{\odot }$ and $M_{\mathrm{ej}%
}=1.3M_{\odot }$ for SNe 1997ef \citep{Iwamoto00} and 2007ru \citep{Sahu09},
respectively. This difference is mainly because in this model the light
curve peak is not caused by $^{56}$Ni. It can be checked that the values
given in Table \ref{tbl:para} is consistent with Equation $\left( \ref%
{eq:tau_m-SN}\right) $ if we realize that $\tau _{m}$ should be set equal to
the rising time of $^{56}$Co contribution in Figures \ref{fig:1997ef} and %
\ref{fig:2007ru}. In these figures the asymptotic expansion velocities are
set to $1.1\times 10^{4}\unit{km}\unit{s}^{-1}$ and $1.3\times 10^{4}\unit{km%
}\unit{s}^{-1}$ for SNe 1997ef and 2007ru, respectively. It is curious that
the derived value $M_{\mathrm{ej}}=7.6M_{\odot }$ by \cite{Iwamoto00} for SN
1997ef is even larger than the value given in this work in spite of the fact
that the rising time in the pure-$^{56}$Ni model is shorter than the rising
time of $^{56}$Co contribution in Figure \ref{fig:1997ef}. We note that the
two models in \cite{Iwamoto00}, i.e. CO60 and CO100, are almost identical in
fitting the light curve of SN 1997ef, but give different values of ejecta
mass, $M_{\mathrm{ej}}=4.6M_{\odot }$ and $M_{\mathrm{ej}}=7.6M_{\odot }$,
respectively. One can check that $M_{\mathrm{ej}}=4.6M_{\odot }$ is
consistent with the light curve rising time if the optical opacity is taken
as $\kappa =0.1\unit{cm}^{2}\unit{g}^{-1}$. This may indicate the
uncertainty in the numerical modeling by \cite{Iwamoto00}.

The $^{56}$Ni yield determined in the pure-$^{56}$Ni model is $M_{\mathrm{Ni}%
}=0.15M_{\odot }$ and $M_{\mathrm{Ni}}=0.4M_{\odot }$ for these two SNe,
respectively. As we said, the yield $M_{\mathrm{Ni}}=0.4M_{\odot }$ is
hardly expected by the magnetar-driven shock $^{56}$Ni synthesis %
\citep{Nishimura15,Suwa15}. In addition, what makes the pure-$^{56}$Ni model
for SN 2007ru more or less unrealistic is the derived ratio $M_{\mathrm{Ni}%
}/M_{\mathrm{ej}}\simeq 0.3$, in tension with the theoretical expectation,
which is predicted to be hardly larger than 0.2 \citep{Umeda08}.

In the middle panels of Figures \ref{fig:1997ef} and \ref{fig:2007ru} we
show the fraction of magnetar rotational energy deposited as the kinetic
energy of the SN ejecta. From these figures it is clear that most ($>99\%$)
of the rotational energy of the magnetars deposits as the kinetic energy of
the SN ejecta and only $<1\%$ of the rotational energy deposits to heat the
ejecta, which is in accord with the expectation \citep{Wang16b}.

To figure out why only $\sim 1\%$ of the rotational energy is enough to heat
the SNe Ic-BL to the observed luminosity, it is beneficial to compare the
rotational energy of the magnetars%
\begin{equation}
E_{p}=2\times 10^{52}I_{45}P_{0,-3}^{-2}\unit{erg}
\end{equation}%
with the decay energy of $^{56}$Ni and $^{56}$Co%
\begin{equation}
E_{\mathrm{decay}}=1.88\times 10^{50}\frac{M_{\mathrm{Ni}}}{M_{\odot }}\unit{%
erg}.
\end{equation}%
Assuming that a fraction $\eta _{E}$ of the rotational energy of the
magnetar $E_{p}$ is converted to the thermal energy of the SN, then the
deposited thermal energy by the magnetar is equivalent to $^{56}$Ni of mass%
\begin{equation}
M_{\mathrm{Ni}}=1.047I_{45}P_{0,-3}^{-2}\eta _{E,-2}M_{\odot },
\end{equation}%
where $\eta _{E}=0.01\eta _{E,-2}$. Because the needed masses of $^{56}$Ni
by the previous analysis are $0.2-0.5M_{\odot }$, it is evident that the
typical initial rotational period of the magnetars that powers the SNe Ic-BL
is $P_{0}\simeq 2\unit{ms}$, in agreement with the values given in Table \ref%
{tbl:para}.

The bottom panels of Figures \ref{fig:1997ef} and \ref{fig:2007ru} show the
evolution of the SN internal energy with initial value $10^{50}\unit{erg}$.
As expected, the SN internal energy increases only slightly despite the
tremendous energy injection rate $\gtrsim 10^{50}\unit{erg}\unit{s}^{-1}$
during the spin-down timescale $\sim 10^{-3}\unit{days}$. This is just why
the SNe Ic-BL powered by a rapidly spinning-down magnetar can gain the
formidable kinetic energy $\sim 10^{52}\unit{erg}$.

With the parameters in Table \ref{tbl:para}, it seems difficult to
understand how a magnetar with spin-down timescale as short as $\sim 10^{-3}%
\unit{days}$ can power an SN lasting for $\sim 20\unit{days}$. This can be
most easily understood by evaluating how much magnetar rotational energy is
left at the light curve peak time $t_{\mathrm{pk}}$. Because $t_{\mathrm{pk}%
}\gg T_{\mathrm{sd}}$, the energy left at time $t_{\mathrm{pk}}$ is%
\begin{equation}
E=E_{0}\left( 1+\frac{t_{\mathrm{pk}}}{T_{\mathrm{sd}}}\right) ^{-1}\sim
10^{48}\unit{erg}E_{0,52}t_{\mathrm{pk},6}^{-1}T_{\mathrm{sd},2}.
\end{equation}%
This energy is just enough for most SNe Ic-BL with peak luminosity $\sim
10^{42}\unit{erg}\unit{s}^{-1}$ lasting for $\sim 10\unit{days}$.

In the analytical model \citep{Wang16b} it is expected that the magnetic
dipole field of the (assumed) magnetars that powers SNe Ic-BL is much
stronger than that powers SLSNe. This is evident from Table \ref{tbl:para}
that the best-fit magnetic dipole field of the magnetar is $B_{p}\sim 10^{16}%
\unit{G}$, whereas the magnetic field in the case of SLSNe is typically $%
B_{p}\sim 10^{14}\unit{G}$.

Finally, because of the high magnetic field strength $B_{p}\sim 10^{16}\unit{%
G}$ and the rapid spinning of the magnetar $P_{0}\simeq 2\unit{ms}$, the $%
^{56}$Ni with mass as low as $0.06M_{\odot }$ can be synthesized by the
magnetar model \citep{Suwa15}. It is clear from above analyses that a
self-consistent magnetar model for the SNe Ic-BL is established.

\section{Discussion}

\label{sec:dis}

Since their discovery, SNe Ic-BL pose an immediate challenge to the
classical SN light curve modeling because the later failed to simultaneously
reproduce the light curve around peak and the late-time linear decline. As a
plausible attempt, \cite{Maeda03} proposed a two-component model for SNe
Ic-BL in which the bright peak is produced by the fast-moving outer
component while the linear tail is attributed to the slower dense inner
component. Because the two-component model and the model presented here both
assume that the linear tail of the SNe Ic-BL light curve can be attributed
to $^{56}$Co decay, it would be beneficial to compare the $^{56}$Ni mass
inferred here with the inner component $^{56}$Ni mass inferred in the
two-component model. For SN 1997ef, \cite{Maeda03} gave the $^{56}$Ni mass
of the inner component $M_{\mathrm{Ni,inner}}=0.08M_{\odot }$, which is
close to our determination taking into account the different value of $%
\kappa $ adopted in these two works. We also note that for the
GRB-associated SN 1998bw, \cite{Maeda03} found $M_{\mathrm{Ni,inner}%
}=0.1M_{\odot }$, which is much smaller than the usually assumed value $M_{%
\mathrm{Ni}}\simeq 0.5M_{\odot }$. This justifies our finding that the $%
^{56} $Ni mass for the tail modeling is usually much smaller than the peak
modeling.

Our results have immediate stimulations for further research. First,
although here we have studied the SNe Ic-BL not associated with GRBs, the
main conclusion can be equally applied to GRB-SNe. It is usually believed
that the central engine of GRBs are black holes %
\citep{Popham99,Narayan01,Kohri02,Liu07,Song16} or magnetars %
\citep{Usov92,Dai98a,Dai98b,ZhangD08,ZhangD09,ZhangD10,Giacomazzo13,Giacomazzo15}%
. However, since it is currently infeasible to identify the GRB central
engine directly because of the cosmological distance scales of GRBs %
\citep{Kumar15}, the researchers instead pursue indirect signatures of black
holes \citep{Geng13,Wu13,Yu15a} and magnetars %
\citep{Dai06,Gao13a,WangLJ13,zhang13,Yu13,Metzger14,Wang15,Wang16a,Li16,Liu16}
that powers the energetic GRBs. Growing indirect observational evidence
suggests that magnetars could act as the central engine of both LGRBs and
SGRBs %
\citep{Dai06,Rowlinson10,Rowlinson13,Dai12,WangLJ13,Wu14,Gao15,Greiner15}.
However, because of the high mass of $^{56}$Ni needed to heat the GRB-SNe,
magnetars are doubted as the candidate central engine of GRBs. With our
demonstration that this high mass of $^{56}$Ni is actually not the case,
such a concern is removed.\footnote{%
We note that \cite{CanoJohansson16} drew the conclusion that GRB-SNe are
powered by $^{56}$Ni decay under the assumption that the central engine of
GRB-SNe is a magnetar. In drawing this conclusion, \cite{CanoJohansson16}
assume that the mangetar's rotational energy is equally divided between GRB
afterglow and SN. This assumption is somewhat unjustified. At the least,
although the jet launching could be the result of the magnetar spin-down, it
seems more likely to be the result of accretion onto the magnetar %
\citep{ZhangD08,ZhangD09,ZhangD10}. Furthermore, \cite{CanoJohansson16} do
not consider the origin of the kinetic energy of the GRB-SNe in their model.
If we accept the assumption that the huge amount of kinetic energy of the
GRB-SNe comes from the rotational energy of the magnetar, the initial
rotational period of the magnetar cannot be as long as given by \cite%
{CanoJohansson16}. Finally, as we mentioned above, our conclusion that the $%
^{56}$Ni mass for the tail modeling is usually much smaller than the peak
modeling is consistent with the finding by \cite{Maeda03}.}

Second, it is still debated how the jet is launched by a rapidly rotating
magnetar. Comparison of SLSNe and SN Ic-BL may have some implications for
this open issue. Observation shows little evidence of jet associated with
SLSNe \citep{Leloudas15}, while some SNe Ic-BL are accompanied by GRB jets.
Even for those SNe Ic-BL not associated with GRBs, jets or aspheric
expansion are frequently observed. Because the initial rotational periods of
the magnetars that power SLSNe and SNe Ic-BL are similar, we hypothesize
that the magnetic field of the magnetar might be essential for the jet
launch given the fact that the magnetic field of magnetar that powers SNe
Ic-BL is much stronger than that powers SLSNe. This hypothesis relies on
future numerical simulations. We note, however, that \cite{Greiner15} found
the magnetic dipole field of the (assumed) magnetar powering
GRB111209A/SN2011kl to be only $\left( 6-9\right) \times 10^{14}\unit{G}$,
close to that powering SLSNe. This may indicate that strong magnetic field
is not a necessary condition for jet launch. Nevertheless, the contamination
by GRB afterglow and host galaxy background makes the GRB-SNe light curves
poorly sampled and parameter degeneracy could bias the fitting values, e.g.
the magnetic dipole field $B_{p}$ and the $^{56}$Ni masses.

Third, we find that the typical magnetic field of the (assumed) magnetars
that powers SNe Ic-BL is $10^{16}\unit{G}$, which is two orders of magnitude
stronger than the field of the magnetars that powers SLSNe, despite the fact
that they are both millisecond magnetars and are all formed during the core
collapse of massive progenitors. This implies that the magnetic
amplification mechanisms \citep{Mosta15} could be quite different. This
calls for more elaborated numerical simulations that take into account more
microphysical processes. The dipole field as strong as $10^{16}\unit{G}$ is
rare but achievable in theoretical aspects. It is expected that the collapse
of the iron core of the supernova progenitor first results in a
proto-neutron star (PNS). The differential rotation of PNS could amplify the
toroidal field to $\sim 10^{16}\unit{G}$ and above \citep{Wheeler00}.
Several magnetic field amplification mechanisms could operate, including the
linear amplification \citep{Dai06}, $\alpha $-$\Omega $ dynamo %
\citep{Duncan92,Thompson93}, and magnetorotational instability %
\citep{Balbus98}. The toroidal field could be amplified to $\sim 10^{17}%
\unit{G}$ before the buoyancy effect takes it to emerge from the neutron
star surface \citep{Kluzniak98,Dai06}. It is possible that the emerged
dipole field could be as strong as $\sim 10^{16}\unit{G}$.

In addition, because the \emph{bona fide} $^{56}$Ni yield is much lower than
previously thought, the kinetic energy-$^{56}$Ni mass relation %
\citep{Mazzali13} should be substantially revised. By doing so some new
insights could be unveiled.

\begin{acknowledgements}
We thank the anonymous referee for his/her constructive comments. This work is supported
by the National Basic Research Program (\textquotedblleft 973" Program) of China under Grant
No. 2014CB845800 and the National Natural Science Foundation of China (grant
Nos. U1331202, 11573014, and 11322328). D.X. acknowledges the support of the
One-Hundred-Talent Program from the National Astronomical Observatories,
Chinese Academy of Sciences. X.F.W. was also partially supported by the
Youth Innovation Promotion Association (2011231), and the Strategic Priority
Research Program \textquotedblleft The Emergence of Cosmological
Structure\textquotedblright\ (grant No. XDB09000000) of the Chinese Academy
of Sciences.
\end{acknowledgements}


\begin{thebibliography}{Klu\'{z}niak \& Ruderman(1998)}
\bibitem[Arnett(1980)]{Arnett80} Arnett, W. D. 1980, ApJ, 237, 541

\bibitem[Arnett(1982)]{Arnett82} Arnett, W. D. 1982, ApJ, 253, 785

\bibitem[Arnett \& Fu(1989)]{Arnett89} Arnett, W. D., \& Fu, A. 1989, ApJ,
340, 396

\bibitem[Balbus \& Hawley(1998)]{Balbus98} Balbus, S. A., \& Hawley, J. F.
1998, RvMP, 70, 1

\bibitem[Bethe(1990)]{Bethe90} Bethe, H. A. 1990, RvMP, 62, 801

\bibitem[Cano et al.(2016a)]{CanoJohansson16} Cano, Z., Johansson Andreas,
K. G., \& Maeda, K. 2016a, MNRAS, 457, 2761

\bibitem[Cano et al.(2016b)]{Cano16} Cano, Z., Wang, S. Q., Dai, Z. G., \&
Wu, X. F. 2016b, arXiv:1604.03549

\bibitem[Chatzopoulos et al.(2012)]{Chatzopoulos12} Chatzopoulos, E.,
Wheeler, J. C., \& Vinko, J. 2012, ApJ, 746, 121

\bibitem[Chatzopoulos et al.(2013)]{Chatzopoulos13} Chatzopoulos, E.,
Wheeler, J. C., Vinko, J., et al. 2013, ApJ, 773, 76

\bibitem[Chen et al.(2015)]{Chen15} Chen, T. W., Smartt, S. J., Jerkstrand,
A., et al. 2015, MNRAS, 452, 1567

\bibitem[Dai \& Liu(2012)]{Dai12} Dai, Z. G. \& Liu, R. Y. 2012, ApJ, 759, 58

\bibitem[Dai \& Lu(1998a)]{Dai98a} Dai, Z. G., \& Lu, T. 1998a, A\&A, 333,
L87

\bibitem[Dai \& Lu(1998b)]{Dai98b} Dai, Z. G., \& Lu, T. 1998b, PhRvL, 81,
4301

\bibitem[Dai et al.(2016)]{Dai16} Dai, Z. G., Wang, S. Q., Wang, J. S.,
Wang, L. J., \& Yu, Y. W. 2016, ApJ, 817, 132

\bibitem[Dai et al.(2006)]{Dai06} Dai, Z. G., Wang, X. Y., Wu, X. F. \&
Zhang, B. 2006, Sci, 311, 1127

\bibitem[Duncan \& Thompson(1992)]{Duncan92} Duncan, R. C., \& Thompson, C.
1992, ApJL, 392, L9

\bibitem[Filippenko(1997)]{Filippenko97} Filippenko, A. V. 1997, ARA\&A, 35,
309

\bibitem[Gal-Yam et al.(2009)]{Gal-Yam09} Gal-Yam, A., Mazzali, P., Ofek, E.
O., et al. 2009, Natur, 462, 624

\bibitem[Gao et al.(2015)]{Gao15} Gao, H., Ding, X., Wu, X. F., Dai, Z. G.,
\& Zhang, B. 2015, ApJ, 807, 163

\bibitem[Gao et al.(2013)]{Gao13a} Gao, H., Ding, X., Wu, X. F., Zhang, B.,
\& Dai, Z. G. 2013, ApJ, 771, 86

\bibitem[Geng et al.(2013)]{Geng13} Geng, J. J., Wu, X. F., Huang, Y. F., \&
Yu, Y. B. 2013, ApJ, 779, 28

\bibitem[Giacomazzo \& Perna(2013)]{Giacomazzo13} Giacomazzo, B., \& Perna,
R. 2013, ApJL, 771, L26

\bibitem[Giacomazzo et al.(2015)]{Giacomazzo15} Giacomazzo, B., Zrake, J.,
Duffell, P. C., et al. 2015, ApJ, 809, 39

\bibitem[Greiner et al.(2015)]{Greiner15} Greiner, J., Mazzali, P. A., Kann,
D. A., et al. 2015, Natur, 523, 189

\bibitem[Inserra et al.(2013)]{Inserra13} Inserra, C., Smartt, S. J.,
Jerkstrand, A., et al. 2013, ApJ, 770, 128

\bibitem[Iwamoto et al.(2000)]{Iwamoto00} Iwamoto, K., Nakamura, T., Nomoto,
K., et al. 2000, ApJ, 534, 660

\bibitem[Janka(2012)]{Janka12} Janka, H-T. 2012, ARNPS, 62, 407

\bibitem[Kasen \& Bildsten(2010)]{Kasen10} Kasen, D., \& Bildsten, L. 2010,
ApJ, 717, 245

\bibitem[Kashiyama et al.(2016)]{Kashiyama16} Kashiyama, K., Murase, K.,
Bartos, I., et al. 2016, ApJ, 818, 94

\bibitem[Klu\'{z}niak \& Ruderman(1998)]{Kluzniak98} Klu\'{z}niak, W., \&
Ruderman, M. 1998, ApJL, 505, L113

\bibitem[Kohri \& Mineshige(2002)]{Kohri02} Kohri, K., \& Mineshige, S.
2002, ApJ, 577, 311

\bibitem[Kumar \& Zhang(2015)]{Kumar15} Kumar, P., \& Zhang, B. 2015, PhR,
561, 1

\bibitem[Leloudas et al.(2015)]{Leloudas15} Leloudas, G., Patat, F., Maund,
J. R., et al. 2015, ApJL, 815, L10

\bibitem[Li \& Yu(2016)]{Li16} Li, S. Z., \& Yu, Y. W. 2016, ApJ, 819, 120

\bibitem[Liu et al.(2016)]{Liu16} Liu, L. D., Wang, L. J., \& Dai, Z. G.
2016, A\&A, 592, A92

\bibitem[Liu et al.(2007)]{Liu07} Liu, T., Gu, W. M., Xue, L., \& Lu, J. F.
2007, ApJ, 661, 1025

\bibitem[Maeda et al.(2003)]{Maeda03} Maeda, K., Mazzali, P. A., Deng, J.
S., et al. 2003, ApJ, 593, 931

\bibitem[Mazzali et al.(2014)]{Mazzali14} Mazzali, P. A., McFadyen, A. I.,
Woosley, S. E., Pian, E., \& Tanaka, M. 2014, MNRAS, 443, 67

\bibitem[Mazzali et al.(2013)]{Mazzali13} Mazzali, P. A., Walker, E. S.,
Pian, E., et al. 2013, MNRAS, 432, 2463

\bibitem[Metzger et al.(2015)]{Metzger15} Metzger, B. D., Margalit, B.,
Kasen, D., Quataert, E. 2015, MNRAS, 454, 3311

\bibitem[Metzger \& Piro(2014)]{Metzger14} Metzger, B. D., \& Piro, A. L.
2014, MNRAS, 439, 3916

\bibitem[Milisavljevic et al.(2015)]{Milisavljevic15} Milisavljevic, D.,
Margutti, R., Parrent, J. T., et al. 2015, ApJ, 799, 51

\bibitem[M\"{o}sta et al.(2015)]{Mosta15} M\"{o}sta, P., Ott, C. D., Radice,
D., et al. 2015, Natur, 528, 376

\bibitem[Narayan et al.(2001)]{Narayan01} Narayan, R., Piran, T., \& Kumar,
P. 2001, ApJ, 557, 949

\bibitem[Nicholl et al.(2014)]{Nicholl14} Nicholl, M., Smartt, S. J.,
Jerkstrand, A., et al. 2014, MNRAS, 444, 2096

\bibitem[Nishimura et al.(2015)]{Nishimura15} Nishimura, N., Takiwaki, T.,
\& Thielemann, F.-K. 2015, ApJ, 810, 109

\bibitem[Pignata et al.(2011)]{Pignata11} Pignata, G., Stritzinger, M.,
Soderberg, A., et al. 2011, ApJ, 728, 14

\bibitem[Popham et al.(1999)]{Popham99} Popham, R., Woosley, S. E., \&
Fryer, C. 1999, ApJ, 518, 356

\bibitem[Rowlinson et al.(2013)]{Rowlinson13} Rowlinson, A., O'Brien, P. T.,
Metzger, B. D., et al. 2013, MNRAS, 430, 1061

\bibitem[Rowlinson et al.(2010)]{Rowlinson10} Rowlinson, A., O'Brien, P. T.,
Tanvir, N. R., et al. 2010, MNRAS, 409, 531

\bibitem[Sahu et al.(2009)]{Sahu09} Sahu, D. K., Tanaka, M., Anupama, G. C.,
et al. 2009, ApJ, 697, 676

\bibitem[Song et al.(2016)]{Song16} Song, C. Y., Liu, T., Gu, W. M., Tian,
J. X. 2016, MNRAS, 458, 1921

\bibitem[Suwa \& Tominaga(2015)]{Suwa15} Suwa, Y., \& Tominaga, N. 2015,
MNRAS, 451, 282

\bibitem[Thompson \& Duncan(1993)]{Thompson93} Thompson, C., \& Duncan, R.
C. 1993, ApJ, 408, 194

\bibitem[Thompson et al.(2004)]{Thompson04} Thompson, T. A., Chang, P., \&
Quataert, E. 2004, ApJ, 611, 380

\bibitem[Umeda \& Nomoto(2008)]{Umeda08} Umeda, H., \& Nomoto, K. 2008, ApJ,
673, 1014

\bibitem[Usov(1992)]{Usov92} Usov, V. V. 1992, Natur, 357, 472

\bibitem[Wang \& Dai(2013)]{WangLJ13} Wang, L. J., \& Dai, Z. G. 2013, ApJL,
774, L33

\bibitem[Wang et al.(2015a)]{Wang15} Wang, L. J., Dai, Z. G., \& Yu, Y. W.
2015a, ApJ, 800, 79

\bibitem[Wang et al.(2016a)]{Wang16a} Wang, L. J., Dai, Z. G., Liu, L. D.,
\& Wu, X. F. 2016a, ApJ, 823, 15

\bibitem[Wang et al.(2016b)]{Wang16b} Wang, L. J., Wang, S. Q., Dai, Z. G.,
et al. 2016b, ApJ, 821, 22

\bibitem[Wang et al.(2016c)]{WangLiu16} Wang, S. Q., Liu, L. D., Dai, Z. G.,
Wang, L. J., \& Wu, X. F. 2016c, ApJ, 828, 87

\bibitem[Wang et al.(2015b)]{WangWang15} Wang, S. Q., Wang, L. J., Dai, Z.
G., \& Wu, X. F. 2015b, ApJ, 799, 107

\bibitem[Wang et~al.(2015c)]{WangWang2015b} Wang, S.~Q., Wang, L.~J., Dai,
Z.~G., \& Wu, X.~F. 2015c, ApJ, 807, 147

\bibitem[Wheeler et al.(2000)]{Wheeler00} Wheeler, J. C., Yi, I., H\"{o}%
flich, P., \& Wang, L. 2000, ApJ, 537, 810

\bibitem[Woosley(2010)]{Woosley10} Woosley, S. E. 2010, ApJL, 719, L204

\bibitem[Woosley \& Bloom(2006)]{Woosley06} Woosley, S. E. \& Bloom, J. S.
2006, ARA\&A, 44, 507

\bibitem[Woosley et al.(2002)]{Woosley02} Woosley, S. E., Heger, A., \&
Weaver, T. A. 2002, RvMP, 74 , 1015

\bibitem[Wu et al.(2013)]{Wu13} Wu, X. F., Hou, S. J., \& Lei, W. H. 2013,
ApJL, 767, L36

\bibitem[Wu et al.(2014)]{Wu14} Wu X. F., Gao H., Ding X., Zhang B., Dai Z.
G., Wei J. Y., 2014, ApJL, 781, L10

\bibitem[Yu et al.(2015)]{Yu15a} Yu, Y. B., Wu, X. F., Huang, Y. F., et al.
2015, MNRAS, 446, 3642

\bibitem[Yu et al.(2013)]{Yu13} Yu, Y. W., Zhang, B., \& Gao, H. 2013, ApJL,
776, L40

\bibitem[Zhang(2013)]{zhang13} Zhang, B. 2013, ApJL, 763, L22

\bibitem[Zhang \& Dai(2008)]{ZhangD08} Zhang, D., \& Dai, Z. G. 2008, ApJ,
683, 329

\bibitem[Zhang \& Dai(2009)]{ZhangD09} Zhang, D., \& Dai, Z. G. 2009, ApJ,
703, 461

\bibitem[Zhang \& Dai(2010)]{ZhangD10} Zhang, D., \& Dai, Z. G. 2010, ApJ,
718, 841
\end{thebibliography}
\end{document}